\documentclass[prl,aps,twocolumn,superscriptaddress]{revtex4-1}

\usepackage[dvips]{graphicx}
\usepackage{amssymb,amsfonts,amsmath}
\usepackage{color}
\usepackage{ulem}

\newcommand{\rv}{{\mathbf r}}

\newcommand{\Jv}{{\bf J}}

\newcommand{\Fv}{{\bf F}}
\newcommand{\fv}{{\bf f}}

\begin{document}

\title{Better than Counting: Density Profiles from Force Sampling}

\author{Daniel de las Heras}
\affiliation{Theoretische Physik II, Physikalisches Institut, 
  Universit{\"a}t Bayreuth, D-95440 Bayreuth, Germany}
\author{Matthias Schmidt}
\affiliation{Theoretische Physik II, Physikalisches Institut, 
  Universit{\"a}t Bayreuth, D-95440 Bayreuth, Germany}

\date{30 October 2017, revised version: 31 January 2018}

\begin{abstract}
Calculating one-body density profiles in equilibrium via
particle-based simulation methods involves counting of events of
particle occurrences at (histogram-resolved) space points.  Here we
investigate an alternative method based on a histogram of the
local force density. Via an exact sum rule the density profile is
obtained with a simple spatial integration. The method circumvents the
inherent ideal gas fluctuations. We have tested the method in Monte
Carlo, Brownian Dynamics and Molecular Dynamics simulations. The
results carry a statistical uncertainty smaller than that of the
standard, counting, method, reducing therefore the computation time.
\end{abstract}

\maketitle

The microscopic one-body density distribution $\rho({\bf r})$ is
arguably the most important order parameter in simple fluids. While in
homogeneous bulk fluid states $\rho=\rm const$, in crystals the
density ``profile'' is peaked at the lattice sites. There is a
multitude of physically interesting situations where the density
$\rho\neq \rm const$, such as for fluids in capillaries, 
across interfaces, under the action of external fields, etc. 
Accurate measurements of $\rho(\rv)$ are very valuable e.g. in order to
study wetting properties~\cite{PhysRevLett.59.98,PhysRevA.39.6402}, capillary effects~\cite{capi}, and
crystal nucleation~\cite{PhysRevLett.91.015703} on substrates,
to characterize the intrinsic liquid-vapor interface~\cite{PhysRevLett.91.166103} and
out-of-equilibrium phase coexistence~\cite{PhysRevLett.90.086601}, 
to determine the charge distributions in capacitors~\cite{capa1,capa2}, 
and  the superadiabatic forces in Brownian systems~\cite{PRLsuperad},
as well as to obtain information about the bulk phase behaviour in
sedimentation-diffusion-equilibrium~\cite{0295-5075-28-9-009,PhysRevE.75.041405,PhysRevE.93.030601}.
Furthermore, within density functional theory (DFT)~\cite{mermin,evans79}, the one-body
density attains a fundamental role in the Mermin-Evans extremal
principle that determines all thermodynamic and structural properties
of the system. High quality simulation data is necessary for the development
and assessment of modern DFT approximations~\cite{roth}.

Experimentally, $\rho(\rv)$ is accessible by a multitude of
methods. Examples in colloidal systems are the analysis of confocal
microscopy data~\cite{confocal1,confocal2}, total internal
reflection microscopy near substrates~\cite{tirm}, and turbidity
measurements~\cite{piazza}. In molecular systems $\rho(\rv)$ can be measured via
three-dimensional AFM scanning \cite{tarazonaExp}.

Mathematically, the one-body density distribution is defined as
\begin{align}
  \rho({\bf r}) &= \left\langle \sum_i \delta({\bf r}-{\bf r}_i)\right\rangle,
  \label{EQdensityDistribution}
\end{align}
where $\rv$ indicates the spatial argument, the sum runs over all
particles, $\delta(\cdot)$ indicates the Dirac distribution, ${\bf
  r}_i$ is the position of particle $i$ and the angles denote the
statistical average, which in equilibrium is carried out over
the appropriate (e.g.\ canonical) ensemble. 

The standard particle-based approach to sample $\rho(\rv)$ is to
discretize the Dirac function and to count events in a histogram,
labelled by position $\bf r$ and with bins of a certain size $\Delta
V$. Normalization by $\Delta V$ and by the number of sampling sweeps
ensures the correct normalization, $\int d\rv \rho(\rv)=N$, where $N$ is
the total number of particles. For cases of additional symmetry, such
as e.g.\ planar problems between, say, parallel walls, the density
profile might depend only on a reduced number of coordinates, say
$\rho(z)$ where $z$ is the coordinate perpendicular to the walls.
In practice brute force can be required to obtain accurate data.

We investigate here an alternative method to sample $\rho(\rv)$, based
on a histogram of the local force density. Working on the level of
force densities has been suggested before. In particular, Borgis
{\it et al}~\cite{Borgis2013}, following the ideas of Assaraf {\it
et al}~\cite{PhysRevE.75.035701} for quantum systems, describe a
variety of advanced methods for sampling the pair distribution
function and the one-body density profile in classical systems. The
approach described in Ref.~\cite{Borgis2013} is based on expressing
formally the delta function in \eqref{EQdensityDistribution} in a
mathematically analogous way to the treatment of a point charge in
electrostatics. This allows to sample distribution functions that
correspond to the electrostatic potential in the analogy. (The
systems considered do not need to carry actual charges for their
approach to work.) The authors find a reduced variance in the
results that they obtain for the distribution functions of interest.
Furthermore the force density plays a central role in advanced
(adaptive resolution) Molecular Dynamics methods, as exemplified by
the work of Fritsch {\it et al}~\cite{PhysRevLett.108.170602}. 
The force density is also relevant for investigations of the potential of mean force,
although in the current work, we do not specify any particular
(reaction) coordinate, as is typically done in characterizing
complex systems by a coarse-grained potential of mean force.

We work on the level of the equilibrium force density balance~\cite{Hansen13}
\begin{align} 
\Fv(\rv) -  k_{\text B}T \nabla \rho(\rv) &= 0,
 \label{EQforceBalance}
\end{align}
where the total (deterministic) one-body force density distribution is given by
\begin{align}
  \Fv({\bf r}) &= \left\langle \sum_i \fv_i(\rv^N )
  \delta({\bf r}-{\bf r}_i)\right\rangle,
  \label{EQforceDensity}
\end{align}
with the total force acting on particle $i$ being
\begin{align}
  \fv_i(\rv^N) &= 
  -\nabla_i u(\rv^N) - \nabla_i V_{\rm ext}(\rv_i), 
\end{align}
where $\nabla_i$ is the derivative with respect to $\rv_i$, $u(\rv^N)$
is the interparticle interaction potential
($\rv^N=\rv_1\ldots\rv_N$), and $V_{\rm ext}(\rv)$ is the external
potential. In Eq.~\eqref{EQforceBalance}, $k_{\text B}$ is the Boltzmann constant, and
$T$ is temperature. 

In short, having sampled $\Fv(\rv)$ allows us to integrate
\eqref{EQforceBalance} in space in order to obtain results for
$\rho(\rv)$. In particular, for effectively one-dimensional problems, 
carrying out a simple one-dimensional integration (along $z$) is all
that is required. In the general case, a line integral needs to be performed,
\begin{align}
  \rho(\rv) &= \rho_0  + (k_{\text B}T)^{-1} \int_{\Gamma} d{\bf s}\cdot \Fv(\rv),\label{line}
\end{align}
where $\Gamma$ represents an appropriate path that connects (say) the
origin with position $\rv$, and $d\bf s$ is the differential line
element. The integral in Eq.~\eqref{line} determines the density profile up to an additive
constant, $\rho_0$, that can be determined by imposing the correct normalization.

Alternatively, we can invert Eq.~\eqref{EQforceBalance}
\begin{equation}
\rho(\rv)=\rho_0+(k_{\text B}T)^{-1}\nabla^{-1}\cdot\Fv(\rv),\label{EQinverseNabla}
\end{equation}
applying the inverse operator $\nabla^{-1}$ to the force density field
\begin{equation}
\nabla^{-1}\cdot\Fv(\rv)=\frac1{c_d}\int d\rv'\frac{\rv-\rv'}{|\rv-\rv'|^d}\cdot\Fv(\rv'),
\end{equation}
where $d$ indicates the dimensionality of the system and the constant $c_d=4\pi$ if $d=3$ and $c_d=2\pi$ if
$d=2$~\footnote{Note that in Ref.~\cite{Borgis2013}
the authors suggest in words the inversion (6) below their Eq.(22),
using Fourier space to solve (7).
}.

The advantage of the force sampling method is that it only samples the (non-trivial)
interaction contribution \eqref{EQforceDensity}. The (ideal gas)
diffusive term, $-k_{\text B}T\nabla\rho$, is treated explicitly. This is in contrast to
sampling $\rho(\rv)$ directly via Eq.~\eqref{EQdensityDistribution}, where these trivial
fluctuations induce a very significant fluctuating background which
besets the data.

To illustrate the accuracy of the new method we carry out
Monte Carlo (MC), Brownian Dynamics (BD), and Molecular Dynamics (MD)
simulations. We compare the density profiles obtained via the 
traditional counting method and the force balance sampling.

In order to have the possibility to provide quasi-exact data, against which to gauge both methods,
we study a system with $N=25$ particles interacting via the Lennard-Jones (LJ) $6-12$ potential.
Hence, the interparticle potential between two particles separated by a distance
$r$ is
$\phi(r)=4\epsilon\left[\left(\frac\sigma r\right)^{12}-\left(\frac\sigma r\right)^{6}\right].$
We set $\epsilon=1$ and $\sigma=1$ as the units of energy and length, respectively. The particles
are located in a square box of side length $L=10\sigma$ with periodic boundary conditions.
The particles are in equilibrium in an external potential
$V_{\text{ext}}(x)=V_0\sin(2\pi n_wx/L)$, that depends only on the $x$-coordinate. We fix $V_0/\epsilon=0.01$ and 
$n_w=5$. The temperature is $k_{\text B}T/\epsilon=1$. We impose a relatively small external potential.
such that the resulting equilibrium $\rho(x)$ is rather flat. The profile shows peak-to-peak oscillations of $\sim 1\%$
relative to the average density.
Sampling such small differences in the density profile is highly demanding and therefore constitutes a strong test
for the force sampling method.  A schematic of the system and plots of both $\rho$ and $V_{\text{ext}}$
are shown in the Supplemental material~\cite{supp}.

\begin{figure*}
\includegraphics[width=0.90\textwidth]{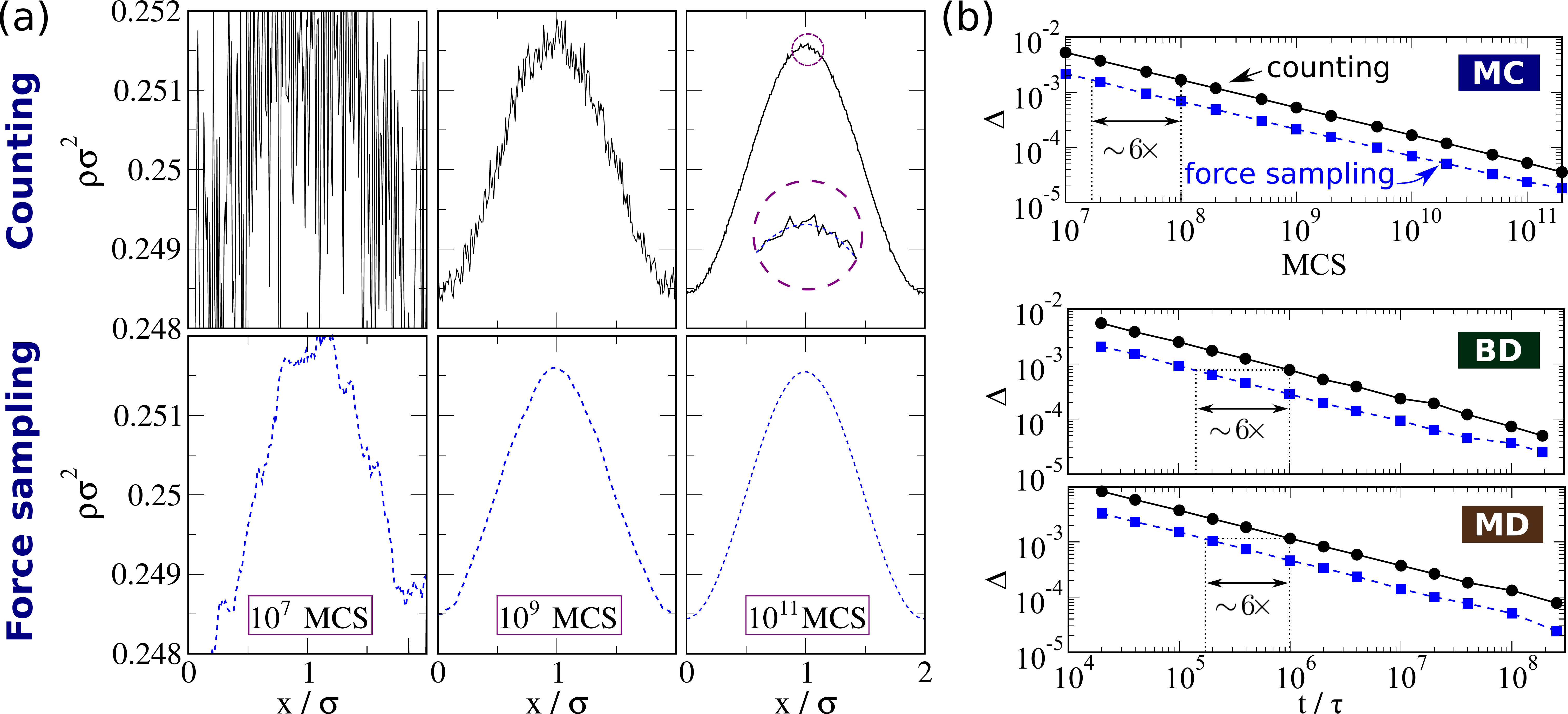}
\caption{(a) Density profiles obtained with MC simulations for different numbers of MCS, as indicated.
The bin size is $\Delta x/\sigma=0.01$, $N=25$, and $L/\sigma=10$. The top panels show $\rho(x)$ obtained via
the traditional counting method (black-solid lines). The density
profiles obtained via force sampling are represented in the bottom panels (blue-dashed lines).
The inset in the top panel with $10^{11}$ MCS is a close
view of both methods in the vicinity of the density peak. Only one fifth of the simulation box, $x/\sigma\in[0,2]$,
corresponding to one density peak is  represented. (b) Logarithmic plots of the 
sampling error $\Delta$ as a function of: (i) the number of Monte Carlo steps in MC simulations (top),
(ii) the simulation time $t/\tau$ in BD simulations (middle) and MD simulations (bottom).
In BD the time is measure in units of $\tau=\sigma^2\gamma/\epsilon$ with $\gamma=1$ the friction coefficient.
In MD, $\tau=\sigma\sqrt{m/\epsilon}$ with $m=1$ the mass of the particles. Data
obtained via counting (black circles) and via force sampling (blue squares).}
\label{fig1}
\end{figure*}

In Fig.~\ref{fig1}a we compare density profiles obtained via counting and force sampling in MC simulations
with a number of Monte Carlo steps (MCS) ranging from $10^7$ to $10^{11}$. In a MCS each particle is once attempted to be moved.
The statistical noise is significantly smaller in the density profiles  obtained via force sampling. Even
after $10^{11}$ MCS density fluctuations are still far from negligible
when using the traditional counting method. We have  obtained  similar differences between both methods in BD and MD. 

To quantify the accuracy of both methods, we define the sampling error $\Delta$ of the density profile as
\begin{equation}
\Delta=\frac{\int d\rv \left|\rho_{\text s}(\rv)-\rho_{\text{eq}}(\rv)\right|}{\int d\rv\rho_{\text{eq}}(\rv)}.
\label{EQerror}
\end{equation}
Here $\rho_{\text s}$ is the sampled density profile and $\rho_{\text{eq}}$ represents the "true" equilibrium profile.
An accurate estimation of $\rho_{\text{eq}}$ is obtained by running a very long MC simulation ($10^{12}$ MCS) and defining $\rho_{\text{eq}}$ as the average 
profile obtained with both methods, counting and force sampling.

Fig.~\ref{fig1}b shows the sampling error in MC, BD, and MD simulations. In all cases force sampling performs significantly 
better than the traditional counting method. To achieve a given sampling error $\Delta$ with traditional counting we need simulations $\sim 6$ times
longer than using force sampling. In other words, force sampling reduces the computation time by $\sim 80\%$.

\begin{figure}
\includegraphics[width=1.00\columnwidth]{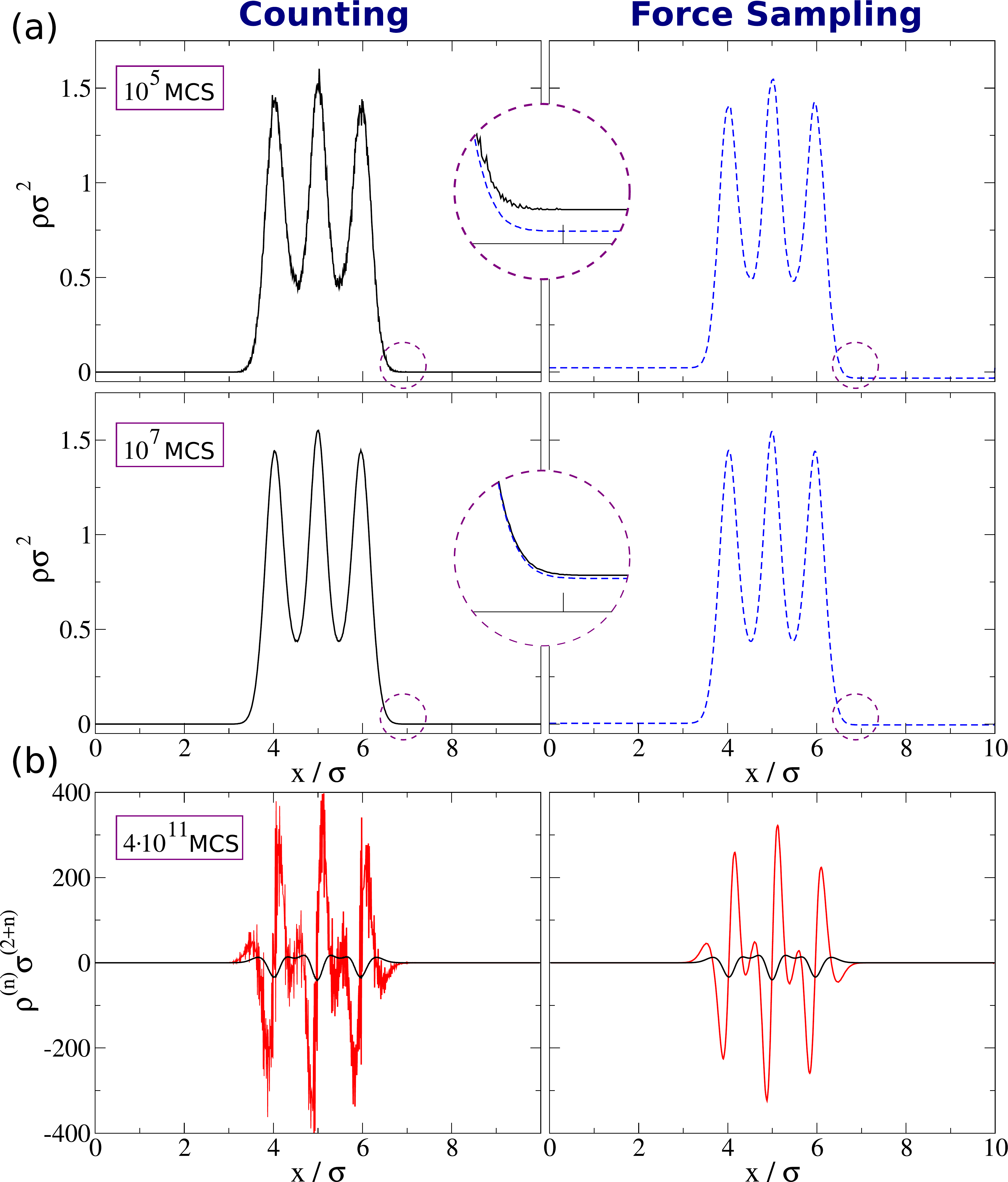}
\caption{(a) Density profiles obtained with MC simulations using counting (left) and force sampling (right).
The number of MCS is $10^5$ (top panels) and $10^7$ (bottom panels), as indicated. The bin size is $\Delta x/\sigma=0.01$.
The particles are in equilibrium in an external 
parabolic potential. The insets are close views of the region where the density vanishes. Using force sampling 
the density in this region reaches an artificial negative value of $\sim-2\cdot10^{-2}$ for $10^5$ MCS
and $\sim-4\cdot10^{-3}$ for $10^7$ MCS. (b) Second $\rho^{(2)}$ (black lines) and third $\rho^{(3)}$ (red lines) numerical derivatives 
(centered difference) of $\rho(x)$ with respect to $x$, $\rho^{(n)}(x)=\partial^{n}\rho(x)/\partial x^{n}$.
Data obtained via counting (left) and force sampling (right) in a MC simulation with $4\cdot10^{11}$~MCS.}
\label{fig2}
\end{figure}

The traditional counting method ensures by construction the correct normalization of $\rho(\rv)$. This is not true when using force sampling.
Here, an additive constant must be added to normalize the density profile. This constant together with the accumulation of the error in the spatial
integral, cf.~\eqref{line} might introduce small artifacts such as slightly asymmetric density profiles in symmetric systems or negative values of the density. 
To illustrate this effect
we introduce a parabolic external potential $V_{\text{ext}}(x)=V_0(x-L/2)^2$, with $V_0/\epsilon=5$. Given the strength of the potential,
the particles strongly accumulate in a small region around $x=L/2$ and the density vanishes in the rest of the simulation box. Due to the normalization
of $\rho$, force sampling erroneously yields a non-zero density value far from $x=L/2$ (positive for $x\ll L/2$ and negative for $x\gg L/2$), see Fig.~\ref{fig2}a.
This anomalous behaviour introduces a relevant error only if the sampling is clearly insufficient. Nevertheless, one must be aware that even 
a small error might be relevant to the calculation of e.g. free energies.
One can partially alleviate this anomaly 
imposing zero density in the regions where no particles have been detected during the simulation.

If the density varies significantly from minimum to maximum, then the profiles obtained with both methods might at first look almost identical. However, building the
numerical derivatives of $\rho(\rv)$ with respect to the spatial coordinates reveals the higher accuracy of force sampling, see Fig.~\ref{fig2}b.

Another example with local high density values is shown in the Supplemental Material~\cite{supp}. Force sampling 
is more accurate and generate smoother profiles than the counting method also at high densities.  The reduction in the sampling error is,
however, less pronounced than in the case of smooth density profiles. We find that force sampling reduces the computation time
by $\sim 40\%$.

The small system size $N=25$ investigated so far has enabled us to carry out a detailed statistical analysis of the relative performance of the two methods.
Nevertheless, the force sampling method remains useful in systems with more realistic values of $N$.

In the Supplemental Material~\cite{supp} we show a comparison of the density profiles obtained with MC
via counting and force sampling in a system with $N=10^3$ and the same "soft" external potential as in Fig.~\ref{fig1}. 
Force sampling is $\sim 5$ times more accurate than counting.

\begin{figure}
\includegraphics[width=0.70\columnwidth]{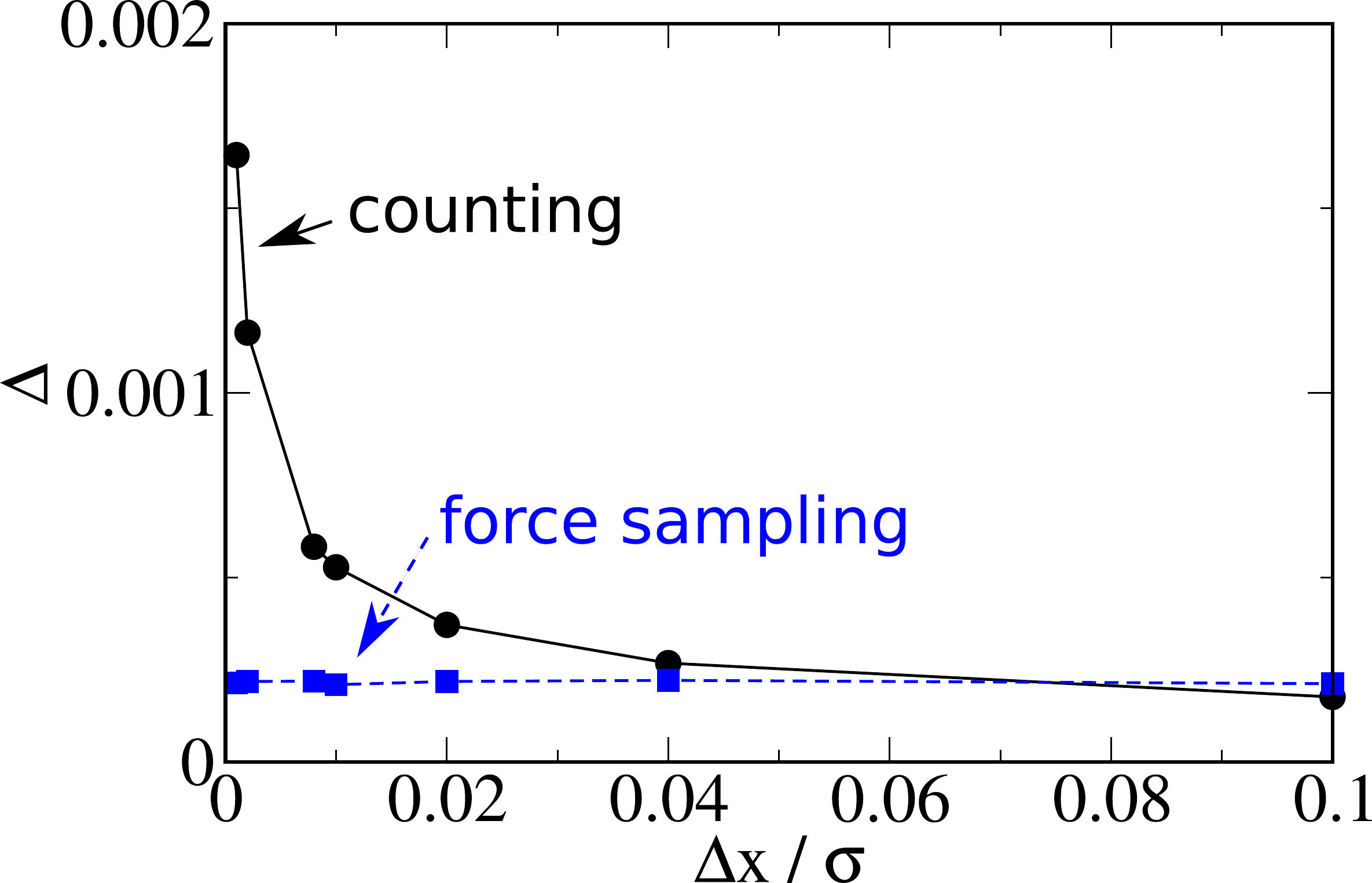}
\caption{
Sampling error $\Delta$ as a function of the bin size. Data
obtained via counting (black circles) and force sampling (blue squares), using MC simulations with $10^9$ MCS.
The "true" equilibrium profile used to compute $\Delta$ is approximated by the average profile of both methods after
$10^{12}$ MCS.}
\label{fig3}
\end{figure}
We discuss next the effect of varying the size of the bin. 
The results shown previously were obtained at a fixed bin size $\Delta x/\sigma=0.01$.
Reducing the size of the bin increases the level of detail with which we can sample the density profile. On the other hand, in the traditional counting
method the error in the density profile also increases by reducing the size of the bin since the number of events contributing to each bin is proportional
to the volume of the bin. In contrast, the error in the force sampling method does not significantly depend on the size of the bin. The reason is that,
in contrast to the counting method, the density at each bin is not determined by the local number of events but via a spatial integral of the force density.
Fig.~\ref{fig3} shows the sampling error as a function of the bin size. The data was obtained using MC simulations
with $10^9$ MCS. The external potential, number of particles, and temperature are the same as those in Fig.~\ref{fig1}.

\begin{figure}
\includegraphics[width=1.00\columnwidth]{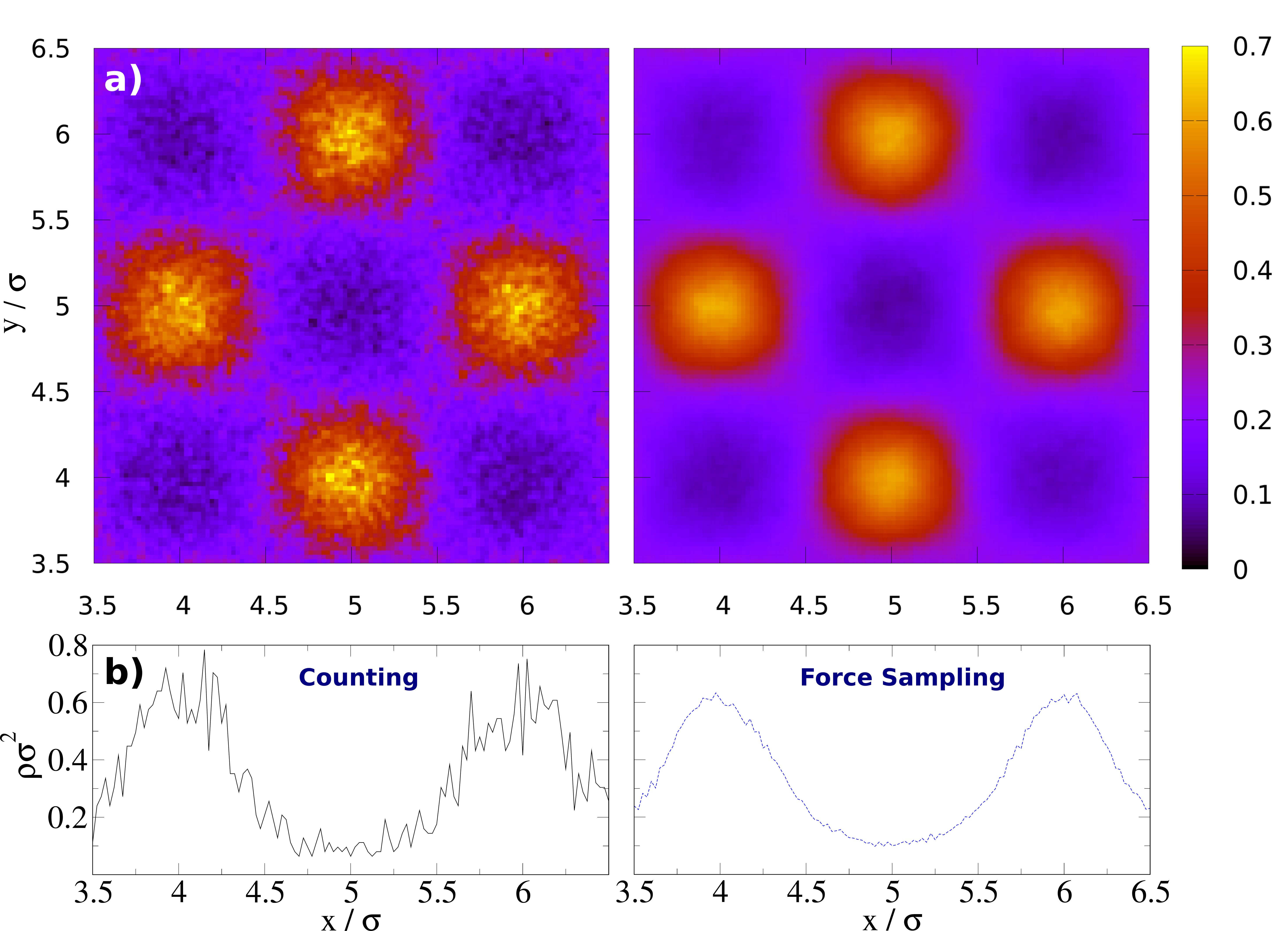}
\caption{
(a) Contour plots of the equilibrium $2$D density profile in the external potential
$V_{\text{ext}}(\rv)=V_0\sin(2\pi n_wx/L)\sin(2\pi n_wy/L)$ with $V_0/\epsilon=1$ and $n_w=5$. Data
obtained with $10^6$ MCS using counting (left) and force sampling (right). The bins are squares of side length $0.025\sigma$.
The simulation box is a square of side length $10\sigma$. Only the central region of the box is shown. (b) Density profile
vs. $x$ at constant $y/\sigma=5$ obtained via counting (left) and force sampling (right).
}
\label{fig4}
\end{figure}

Systems that require a small bin size, such as for example cases where the density profile depends on two or three spatial coordinates,
will substantially benefit from applying the force sampling method. To illustrate this we introduce the external potential
$V_{\text{ext}}(\rv)=V_0\sin(2\pi n_wx/L)\sin(2\pi n_wy/L)$, that depends on both $x$- and $y-$coordinates. We fix $V_0/\epsilon=1$ and
$n_w=5$. The resulting two-dimensional ($2$D) density profile is shown in Fig.~\ref{fig4}. Force sampling is clearly superior to counting.
Details on how to obtain the $2$D density profile using force sampling are given in the Supplemental Material~\cite{supp}.

Implementing the force sampling method is straightforward in all simulation techniques analyzed here and its computational
demand is negligible in both BD and MD, and very low in MC. Note that in MC we need to implement the calculation of the forces (not inherent
in the method) and compute them every certain number of MCS as part of the sampling process.
We sample the force density during the whole simulation, but it is only at the
end of the simulation run that we compute $\rho(\rv)$ via spatial integration of the force density, cf.~\eqref{line}. 

The force sampling method cannot be directly used in hard core systems (i.e., particles interact only if they overlap, in which case 
the interparticle potential is infinity). However, a hard core system can be approximated by a quasihard potential
that decays  very fast with the distance between the particles. 

It might also be possible to extend the force sampling method to study hard core systems using Event Driven Molecular~\cite{EDMD}
and Brownian~\cite{EDBD1,EDBD2} Dynamics since in both cases the moment transfer in a collision is available.

Besides the examples shown here, we have tested the validity and accuracy of force
sampling in a one-dimensional system of quasihard spheres ($\phi(r)\propto r^{-42}$) 
and in one- and two-dimensional systems of Gaussian particles.
In all cases the force sampling method has provided better accuracy than the standard counting method. 

Finally we have also verified the better performance of the method in a $2$D system of Gaussian particles under stationary  shear conditions~\cite{grad_v_3}.
In steady state the method is still valid for those Cartesian components for which Eq.~\eqref{EQforceBalance} still holds, such as e.g. in the direction
perpendicular to the shear flow. Furthermore, in BD it is also possible to apply the force sampling method to out-of-equilibrium conditions,
even away from steady-states. 
In out-of-equilibrium BD the force density balance at any time $t$ is given by
\begin{align} 
\Fv(\rv,t) -  k_{\text B}T \nabla \rho(\rv,t) &= \Jv(\rv,t),
 \label{BDforceBalance}
\end{align}
where $\Jv$ is the one-body current, which in contrast to the equilibrium case, Eq.~\eqref{EQforceBalance}, does not vanish in general.
Therefore in addition to the sampling of the forces it is also necessary
to sample the total current to be able to obtain the density profile via spatial integration. Sampling the current, which is possible using the numerical
derivative of the position vector or via the continuity equation, might increase the statistical noise with respect to the equilibrium case. Therefore, 
it is not guaranteed that force sampling will perform better than counting in out-of-equilibrium. This study constitutes the subject of future work.
Adding inertial terms to Eq.~\eqref{BDforceBalance} would allow the extension of the method to out-of-equilibrium MD.

The generalization of the method to multicomponent mixtures is  straightforward. Testing the performance in grand canonical MC schemes~\cite{capi} 
is an interesting research task for the future.

\begin{acknowledgments}
We thank N.~B. Wilding, E. Chac{\'o}n, M. Dijkstra, M.P. Allen, and D. Borgis for feedback and stimulating discussions, and D.~Borgis for pointing out Ref.~\cite{Borgis2013} to us.
\end{acknowledgments}

\section{Supplemental Material: Better than Counting: Density Profiles from Force Sampling}

\subsection{System}
A schematic of the system is shown in Supplemental Fig.~\ref{figs1}a. The external potential and the corresponding equilibrium density profile are shown in Supplemental Fig.~\ref{figs1}b and c, respectively.
\begin{figure}
\includegraphics[width=0.90\columnwidth]{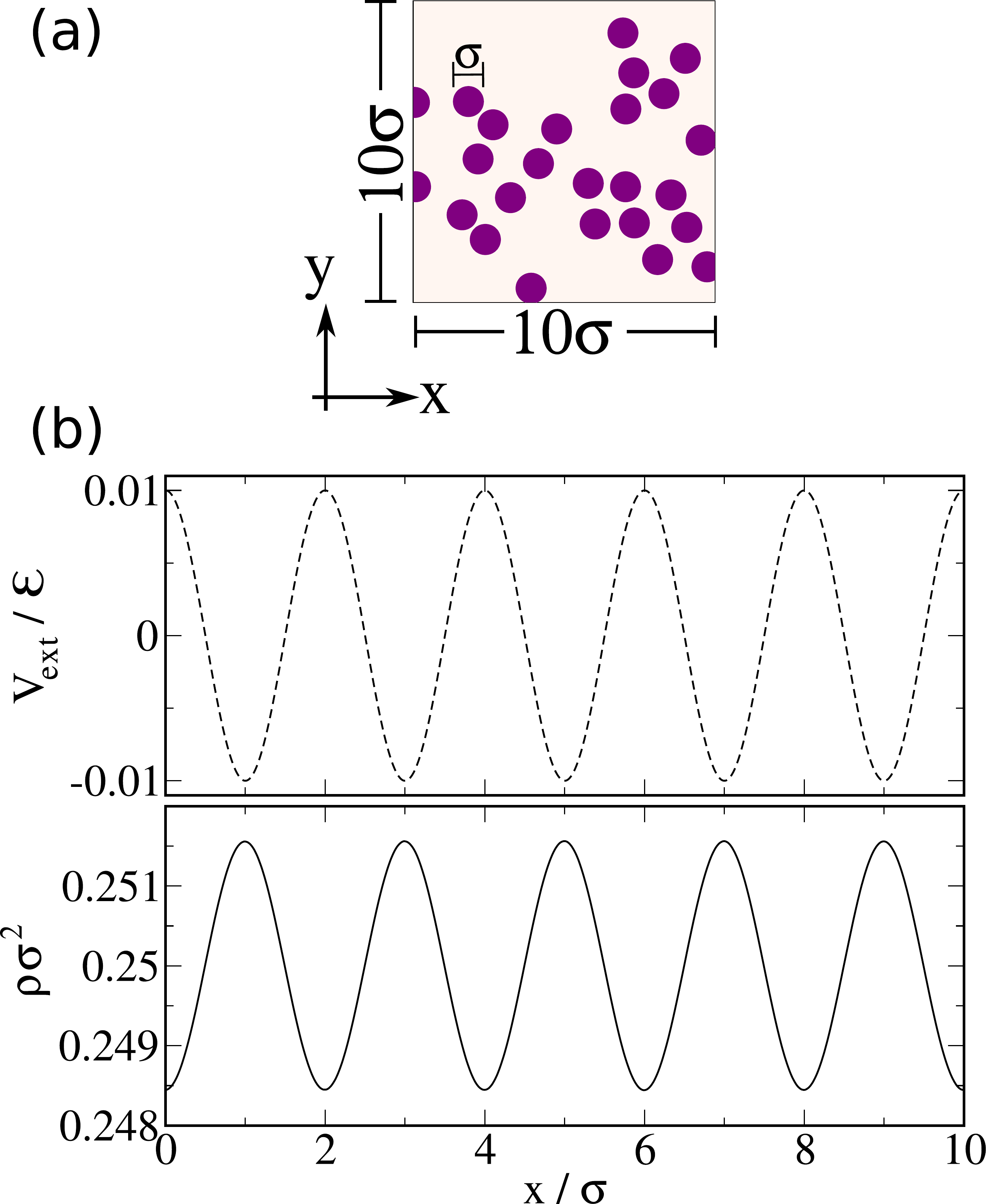}
\caption{(a) Schematic of the system, $N~=~25$ LJ particles of size $\sigma$
in a square box of side length $10\sigma$ with periodic boundary conditions. 
(b) External potential (top)
and corresponding equilibrium density profile (bottom) obtained with MC simulation ($10^{12}$ Monte Carlo steps).}
\label{figs1}
\end{figure}

\subsection{High density states}

An example exhibiting local high density values is shown in Supplemental Fig.~\ref{fig-HD}. The particles are in equilibrium in the same type of external potential as that shown in Fig.~1b
of the main text, that
is  $V_{\text{ext}}(x)=V_0\sin(2\pi n_wx/L)$ with $n_w=5$, but much stronger,  $V_0 / \epsilon=3$. As a result the density profile shows strong oscillations, see Supplemental Fig.~\ref{fig-HD}a. 
The force sampling method is more accurate and generates smoother profiles than the counting method. 
The reduction in the sampling error is, however, less pronounced than in cases with smooth density profiles.
To achieve a given sampling error $\Delta$ with traditional counting we need simulations $\sim 2$ times longer than using force sampling, see 
Supplemental Fig.~\ref{fig-HD}b. That is, force sampling reduces the computation time by $\sim 40\%$.

\begin{figure}
\includegraphics[width=0.80\columnwidth]{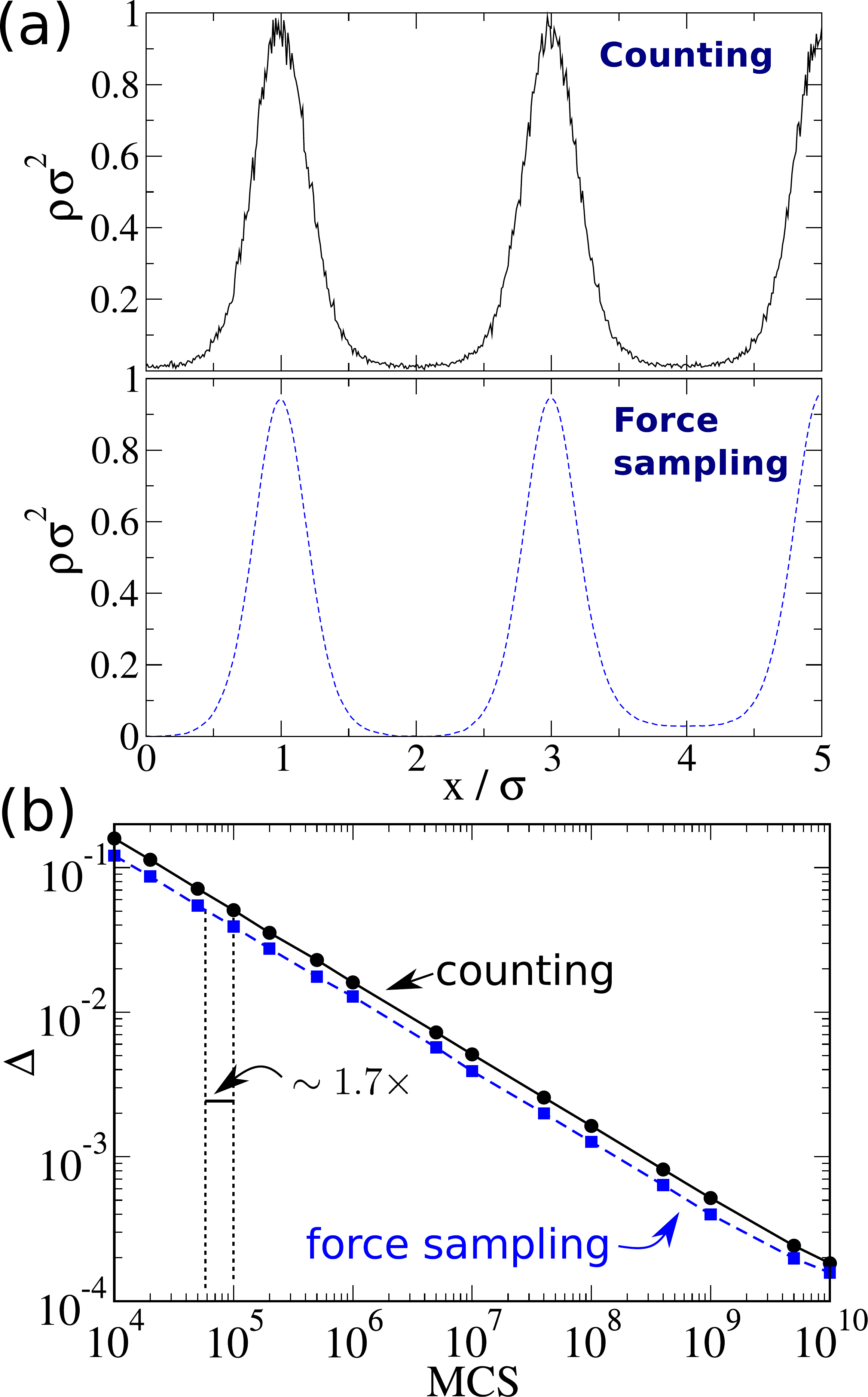}
\caption{
(a) Density profiles obtained with MC simulations using the counting (top) and the force sampling (bottom) method.
The number of MCS is $10^4$ and the bin size is $\Delta x/\sigma=0.01$.  The size of the box is $L/\sigma=10$ (only half of the box is shown) and
$N=25$. The particles are in equilibrium in the external potential $V_{\text{ext}}(x)=V_0\sin(2\pi n_wx/L)$ with $n_w=5$ and $V_0 / \epsilon=3$.
The temperature is $k_{\text B}T/\epsilon=1$.
(b) Logarithmic plot of the sampling error $\Delta$ as a function of the number of Monte Carlo steps. Data
obtained via counting (black circles) and via force sampling (blue squares).
The "true" equilibrium profile used to compute $\Delta$ is approximated by the average profile given by both methods after
$4\cdot10^{11}$ MCS.}
\label{fig-HD}
\end{figure}

\subsection{Realistic number of particles}

\begin{figure}
\includegraphics[width=0.80\columnwidth]{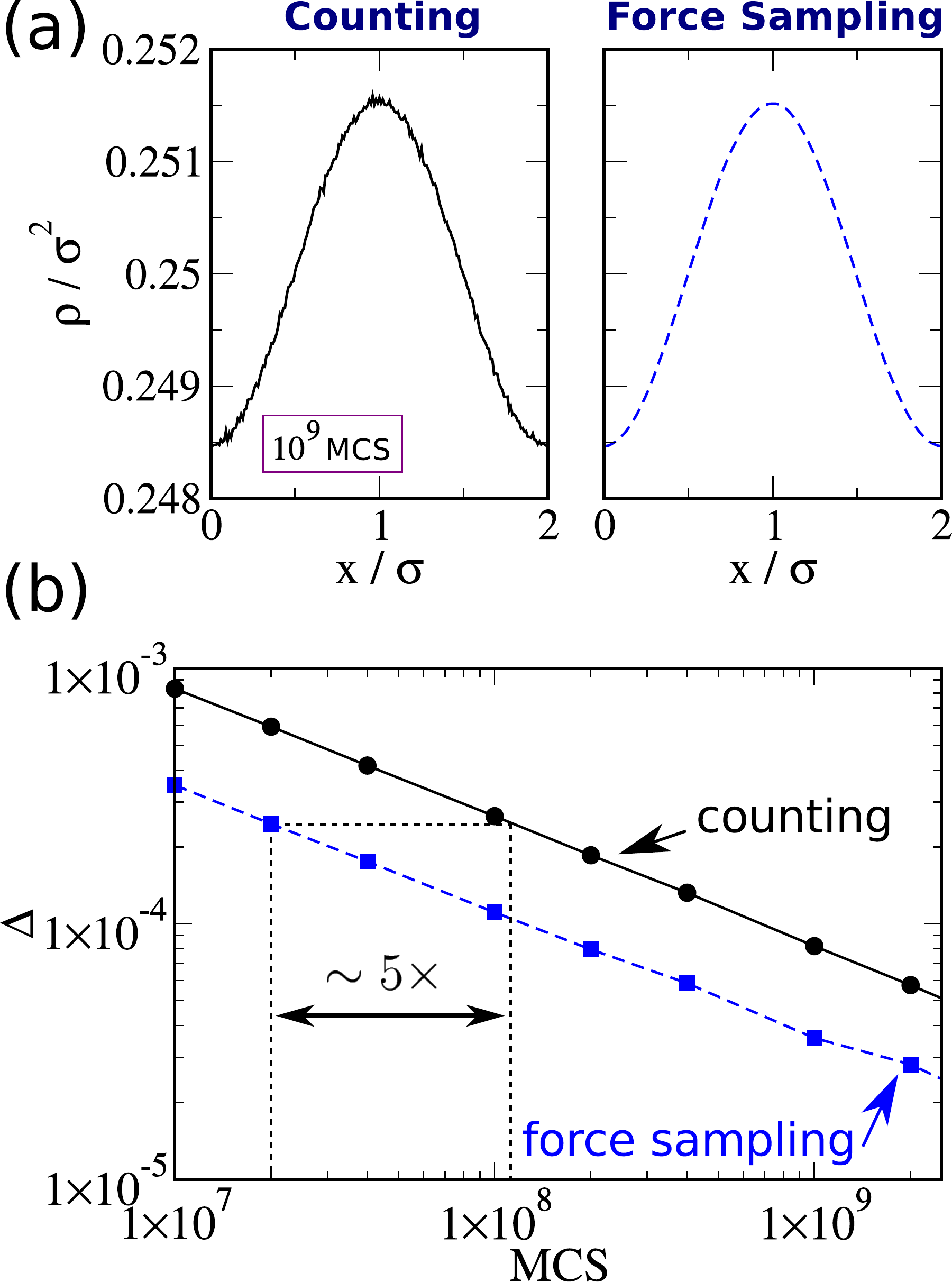}
\caption{(a) Density profiles (MC simulation with $10^9$ MCS) obtained via counting (left) and force sampling (right) 
in a system with $N=10^3$ confined in a box with side lengths $L_x/\sigma=10$ and $L_y/\sigma=400$. The bin size is $\Delta x/\sigma=0.01$.  
The particles are in equilibrium in the external potential $V_{\text{ext}}(x)=V_0\sin(2\pi n_wx/L_x)$, with $V_0/\epsilon=0.01$ and
$n_w=5$. Only one fifth of the simulation box is shown, $x/\sigma\in[0,2]$. (b) Logarithmic plot of the sampling error as a function 
of the number of MCS.}
\label{figs4}
\end{figure}

In Supplemental Fig.~\ref{figs4}a we show a comparison of the density profiles obtained with MC
via counting and force sampling a system with $N=10^3$. The particles are in a rectangular box with side lengths $L_x/\sigma=10$ and $L_y/\sigma=400$
subject to the external potential shown in Fig.~\ref{figs1}b. Therefore, the system is homogeneous in the $y-$coordinate.
In Supplemental Fig.~\ref{figs4}b we show the sampling error $\Delta$ of both methods in MC.
Force sampling is $\sim 5$ times more accurate than counting. The "true" equilibrium profile
used to compute the sampling error $\Delta$ is approximated here by the average profile given by both methods after 
$2\cdot10^{10}$ MCS (obtained by averaging $2\cdot10^{3}$ MC simulations of $10^7$ MCS each).\\

\subsection{Multidimensional density profiles}
If the density profile depends on several spatial coordinates, there are at least three routes to implement the force sampling method. One possibility,
as described in Eq. (5) of the main text, is to perform a line integral of the force density. Alternatively, we can invert the force density balance equation and
obtain the density profile via a volume integral over the full space, see Eqs. (6) and (7) of the main text. Eq. (7) of the main text can be solved either in real
or in Fourier space (see Ref. [21] of the main text) a post processing of the  data sampled during the simulation. Finally, we can also obtain the density profile
via numerical minimization of the functional
\begin{equation}
H[\rho]=\int d\rv||\nabla\rho(\rv)-\Fv(\rv)||^2,
\end{equation}
which is a standard procedure to numerically find the scalar potential that generates a given curl-free vector field. In practice, inverting the force balance equation via Eq. (7) of the
main text, or numerically minimizing $H[\rho]$ results in more accurate density profile than solving the line integral in Eq. (5) of the main text. Note that both inversion of the force
balance equation and minimization of $H[\rho]$ use information from the whole system in order to compute 
the local density profile at position $\rv$.
The two-dimensional density profiles shown in Fig. 4 (right) of the main text have been obtained via minimization of $H[\rho]$.

\end{document}